\documentclass[twocolumn,10pt]{article}
\usepackage[utf8]{inputenc}

\usepackage{color}
\usepackage{graphicx}
\usepackage{caption}
\usepackage{url}
\usepackage{hyperref}
\usepackage{breakurl}

\title{Integrating a large-scale testing campaign in the CK framework}
\author{Andrei Lascu, Alastair F. Donaldson}
\date{}

\begin{document}

\maketitle

\begin{abstract}
We consider the problem of conducting large experimental campaigns in programming languages research. Most research efforts require a certain level of bookkeeping of results. This is manageable via quick, on-the-fly infrastructure implementations. However, it becomes a problem for large-scale testing initiatives, especially as the needs of the project evolve along the way. We look at how the Collective Knowledge generalized testing framework can help with such a project and its overall applicability and ease of use. The project in question is an OpenCL compiler testing campaign. We investigate how to use the Collective Knowledge framework to lead the experimental campaign, by providing storage and representation of test cases and their results. We also provide an initial implementation, publicly available.
\end{abstract}

\section{Introduction}
Our experience is that most experimental work in programming languages research employs purpose-built, and to some extent ad-hoc, infrastructure. One such example is the work done in our PLDI'15 paper \cite{clsmith}, where we ran tens of thousands of tests across multiple machines. Aiming to identify bugs in several OpenCL implementations, we produced a large infrastructure to generate random kernels, test them on a specific configuration, record the results and finally analyze them to obtain some macroscopic observations. It served its purpose well, but significant effort was required to maintain the infrastructure and add new features as the needs of the project evolved. Although the infrastructure was robust enough and sufficiently documented for others to reproduce our experiment, we speculate it would not be easy for third parties to use and extend for further experimental work. This is partly due to a lack of fully comprehensive documentation (which we could address), but more fundamentally due to the ``one-shot'' nature of the infrastructure, tailored specifically for our particular project. The Collective Knowledge (CK) framework \cite{collectivemind2, collectivemind} aims to offer a generalized method for running experiments, recording their results and querying properties of these results. This is particularly interesting to us, due to the sheer volume of data we had to handle. In addition, CK also comes with its own HTML-based experiment viewer and data storage capabilities. This unified data representation, alongside the online repository feature CK offers, makes gathering results from a variety of different platforms straightforward. We describe how integrating our testing mechanism in CK compares to our initial self-made implementation, what extra features we got from using CK, and comment on our experience using the framework.

\section{Background}
\subsection{Many-core Compiler Fuzzing}
The aim of our work \cite{clsmith} was to find compiler bugs for the Open Computing Language (OpenCL) \cite{opencl12}. The techniques we used are \emph{Random Differential Testing} (RDT) \cite{csmith} and \emph{Equivalence Modulo Inputs Testing} (EMI) \cite{emi}.

\paragraph{OpenCL} Open Computing Language (OpenCL) \cite{opencl12}, from the Khronos Group, enables parallel code to be executed on a range of multi-core devices, including CPUs and GPUs. A \emph{host} program initialises memory on the device, handles runtime parameters (e.g. inputs, number of parallel threads) and delegates compilation and execution of the code to be executed on the multicore device, known as a \emph{kernel}. The host program is written in C or C++ augmented with API calls to interface with OpenCL. Conversely, the kernel is written in \emph{OpenCL C}, a subset of C99 extended with additional features. The draw of OpenCL is that it specifies how the host and device should interact, without providing an actual compiler. Granted, it would be near impossible to provide a piece of software for every single OpenCL-compatible hardware device. Therefore, the responsibility of providing a compiler is left to the hardware vendor.

\begin{figure}[h]
  \centering
  \captionsetup{justification=centering}
  \caption{Example of random differential testing. A possible bug is detected when compiling with Compiler 1.}
  \label{fig:sketchrdt}
  \includegraphics[scale=0.75]{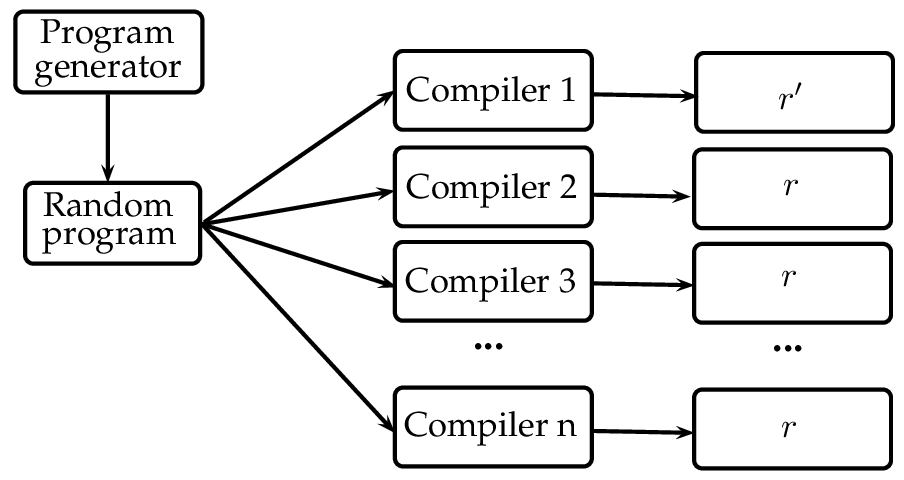}
\end{figure}

\paragraph{Random Differential Testing} As sketched in Figure \ref{fig:sketchrdt}, this technique involves using multiple implementations to compile and execute a given input program. If there is disagreement between implementations regarding the result of the test, at least one of the implementations must exhibit a bug. A requirement of this testing method is that the tests must be well-defined (having no undefined behaviour according to the language specification \cite{opencl12} and agreeing on implementation-defined behaviour) and deterministic (every execution should yield the same result). This technique is used in the Csmith \cite{csmith} tool, which applies it in the C universe. The authors of the tool have found many C bugs in \emph{gcc} and \emph{clang} by generating random, input-free C programs and comparing the results across multiple versions of the compilers.

\begin{figure}[h]
  \centering
  \captionsetup{justification=centering}
  \caption{Example of EMI testing; variant 1 produces a different result than the others, exposing a possible bug.}
  \label{fig:sketchemi}
  \includegraphics[scale=0.75]{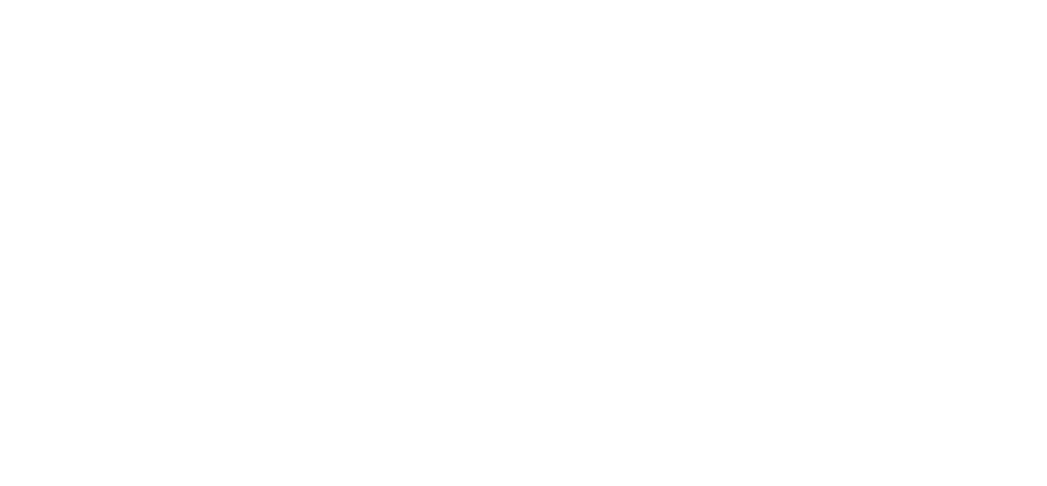}
\end{figure}

\paragraph{Equivalence Modulo Inputs} A more recent testing method, it requires only one program and one implementation (outlined in Figure \ref{fig:sketchemi}. Given a program taking an input, $I$, we profile it to identify any $I$-\emph{dead} code, meaning code that is unreachable under the input $I$. We can modify these sections arbitrarily, without changing how the program behaves on input $I$, only requiring to adhere to the syntactical rules of the programming language. Thus, by mutating the $I$-dead sections, we can obtain multiple variants of the initial program that are functionally identical with respect to input $I$. If a variant thus obtained produces a different result than the initial one, it is certain that there is a bug in the compiler. For our study \cite{clsmith}, we use a slightly modified version of EMI by not looking for $I$-dead sections, but injecting them ourselves. This is due to the lack of a code-coverage profiling tool, according to our knowledge.

\paragraph{CLsmith} We have implemented the above testing methods in a tool, CLsmith,\footnote{\url{https://github.com/ChrisLidbury/CLSmith}} built upon the \emph{Csmith} \cite{csmith} C fuzzer. We lifted the language of the generated random programs from C to OpenCL C and further augmented it with the capability of generating code for certain OpenCL-specific features, which are turned off by default. We selected six combination of features to generate programs with, which we shall call \emph{modes}. We give a brief description of each of the modes (for more detailed information, please refer to Section 4 of \cite{clsmith}):
\begin{itemize}
 \item \textsc{Basic} mode, lifting a Csmith generated program to the OpenCL environment;
 \item \textsc{Vector} mode, testing the \emph{vector} built-in types and operations of OpenCL;
 \item \textsc{Barrier} mode, simulating inter-thread communication;
 \item \textsc{Atomic Section} mode, testing OpenCL atomic operations;
 \item \textsc{Atomic Reduction} mode, a second method for exercising atomic operations;
 \item \textsc{All} mode, combining all of the above.
\end{itemize}

We then began an extensive experimental campaign, spanning 21 OpenCL implementation (i.e.. combination of hardware device and software driver). We ran an initial benchmarking test, where we checked where the implementations lie relative to a reliability threshold. An implementation is deemed to lie below the reliability threshold if, for 600 programs generated by CLsmith, at least 25\% of them lead either to a compiler or runtime crash or a non-majority result. We thus eliminated 11 of our total configurations, as they did not meet this reliability threshold. For the 10 remaining ones, we generated a total of 60000 tests to use via the RDT technique. Due to bugs in CLsmith we discovered later on during the experimental campaign, we had to remove 3185 of these tests due to them being non-deterministic, leaving us with 56,815 programs overall. For EMI testing, we generated 250 \emph{base} programs, which we used to produce 40 variants for each, leading to a total of 10,000 tests. However, 70 of the base programs were later found to be non-deterministic due to a generator bug; removing these and their variants leaves a total of 180 base programs with 7,200 variants.

\paragraph{Experimental analysis} We ran our 56815 tests on all of the 10 platforms and then collected the output of each test. Based on this, we classified each execution (test run on one configuration) as being a compiler crash, a runtime crash, a timeout (we enforced a timeout on execution time), a wrong code bug (we compute a majority result by counting how many instances of one particular result we have; a wrong code bug is when the result of a configuration is different than the computed majority) or a correct execution (same result as the majority result). In the case of an equal split (there is no majority number of configurations giving one single result), we label the test as inconclusive. We gathered all these results to observe weaknesses of certain configurations in certain circumstances. For our full testing campaign, including application of EMI on real-world examples and randomly generated CLsmith tests, please refer to Section 7 of \cite{clsmith}.

\subsection{The Collective Knowledge Framework}
The Collective Knowledge Framework and Repository (CK) \cite{collectivemind2, collectivemind} is a light-weight dependency-free Python library and application. It aims to enable open, collaborative and reproducible research, experimentation, knowledge sharing and predictive analysis.  It also helps in preserving and organising code and experimental data (e.g. benchmarks, tests, libraries, results) via a simple JSON-based API. By abstracting away calls to tools and hardware using the same wrapper API, it allows users to focus on the experimental data at hand, without the need of low-level knowledge. Management of modules is done with a \emph{DOI-style} unique ID (UID) system, allowing components to be indexed by third-party tools and combined into new projects as needed. CK also simplifies connection with predictive analysis tools, such as \emph{scikit-learn} \cite{scikit-learn}, allowing non-specialists to perform statistical analysis of their results.

One of the main aims of CK is to improve collaboration between researchers, by providing necessary tools to re-run previous experiments and recording data that might prove interesting. As this data recording feature is customisable, this would allow other researchers than the original authors of a project to analyse the experimental evaluation from a different perspective. In addition, similar projects could be validated by sharing their respective datasets, provided they are compatible. The CK framework also provides integration with certain third-party software (e.g. IDEs), which could help users feel more comfortable into adopting it.

\section{Implementation}
For the purposes of this paper, we implemented the required tools to evaluate the kernels generated by the CLsmith tool in the CK framework. From the CLsmith project, we took the host-code required to start up the execution of a generated kernel and 600 random tests freshly generated using the latest version of CLsmith (which incorporates some recent bug-fixes). There are two in-depth guides available that describe in detail how to obtain CK and the corresponding CLsmith repository, how to run the available tests and further technical details involved. They are \emph{Getting started with CK} \cite{ckguide} and \emph{Getting started with CLsmith and CK} \cite{ckclsmith}. We now take a detailed look at the process required to run tests and record their results as it was in our initial experimental campaign and how it is with CK integration.

\begin{figure}[h]
  \centering
  \captionsetup{justification=centering}
  \caption{Outline of the components of our initial testing method. The dependencies between the various components made this flow hard to manage.}
  \label{fig:testinginit}
  \includegraphics[scale=0.75]{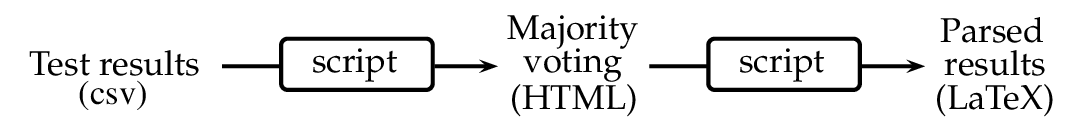}
\end{figure}

\subsection{Initial testing process}
For our paper \cite{clsmith}, we performed testing in six batches corresponding to the CLsmith modes, starting out with generating the required number of tests on a strong, reliable platform (Nvidia GeForce Titan). In order to execute the kernel, we also had to write a host-code application, which we called \texttt{cl\_launcher}. This is a C binary that allocates device resources, compiles and runs the device code (i.e. the kernel) and returns the result. This application was written to run a single test; thus, to run a whole batch of them, we were required to write a separate script to consecutively call \texttt{cl\_launcher} and redirect its output to a csv file. This gives us six output files per configuration, one per mode, which hold the output of every execution. Further, we wrote another script (represented by the first script in Figure \ref{fig:testinginit}) to gather the results of one mode for all platforms into a single HTML table, also calculating the majority result in the process . This allowed us to visually identify programs that would crash or give a different result than the majority in order to further analyse them. Finally, we used one last script (second script box in Figure \ref{fig:testinginit})  to obtain some macro results from these six big tables that allowed us to deduce how well each mode performed at detecting bugs and how platforms compare to one another in terms of reliability, among other results. These tables are Tables 3 and 4 in our original paper \cite{clsmith}. It should be mentioned that these scripts (five in total, as we required two different sets of two scripts for RDT and EMI, as seen in Figure \ref{fig:testinginit}, alongside the one script to continuously execute tests) are not reusable for other work without heavily refactoring, as it specifically depends on the output of the configuration of the pipeline. 

An issue that appeared during our testing initiative was faulty tests. As we analysed our results, we detected some bugs that made certain tests require regeneration. However, as we had already gathered the result data in our csv files, it was very hard to filter only these buggy results out. We ended up mostly having to do redundant work to regenerate fresh results or further update our scripts to handle this particular issue. This meant more time invested in our ad hoc testing infrastructure.

\subsection{Testing via the CK framework}
We now consider using CK for our testing infrastructure, having it control the flow of the testing process, invoking the low-level tools that we provide. In order to integrate our testing into the CK framework, we had to provide only \texttt{cl\_launcher} and some test files. After having this work done by an expert, running tests was as simple as issuing a command line instruction. There was an initial issue that the \emph{stdout} and \emph{stderr} of the application were not recorded by the framework as they were not necessary for previous projects, but that was implemented by the CK developers at our request. We also made use of CK's built-in HTML-based experiment view, which was adapted to record some information we were interested in (e.g. OpenCL version or device used for testing). Another useful feature at our disposal is the ability to re-run individual tests. The experimental view provides the option of selecting which test one would like to re-run and displaying the exact command required, the user only needing to copy it to the command line. Furthermore, the fact that all the results are saved in a customisable framework means that we can adapt it according to our increasing needs as the project develops. This feature is what we find most attractive about the framework.

\begin{figure}[h]
  \centering
  \captionsetup{justification=centering}
  \caption{Outline of testing procedure using CK integration.}
  \label{fig:testingck}
  \includegraphics[scale=0.55]{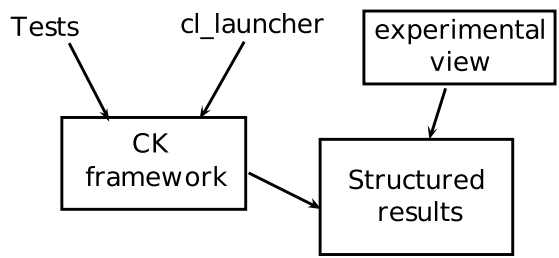}
\end{figure}

\paragraph{CK integration details}
The technical work required to make our tool available in CK was done by Grigori Fursin, a senior developer of the CK framework. The work was done in roughly half a day, the bulk of it being the restructure of our existing data into CK-compliant components. Thus, \texttt{cl\_launcher} became a CK \emph{program} and our 600 tests became \emph{datasets}. These are complemented by two CK \emph{scripts} (JSON files that describe which datasets to run using which program, in addition to multiple CK parameters), which launch 100 existing \textsc{Basic} and \textsc{Vector} mode tests. These containers are not part of CK per se, but have been re-used from existing CK projects. Most of the work was actually describing these new components in JSON format. The final (and, for us, most interesting) element is the \emph{experiment view}. CK contains a \texttt{CK-web} repository, which is able to start a local server giving access to a HTML-based UI containing all the local CK information. We show our view of the entire CK infrastructure for our project in Figure \ref{fig:testingck}. The information presented for CK \emph{experiments} (obtained by running programs on datasets via scripts) can be customised via the aforementioned experiment views. This selects which records to be displayed from all the recorded information. This means users could have multiple perspectives when looking at the same set of results.

However, this data has to be recorded when the experiments are run and is, most likely, individual to each project in part. In our project, this is done via \emph{xOpenME}, a variant of the \emph{OpenME}\footnote{\url{https://github.com/ctuning/openme}} plugin, an event framework that can expose various internal parameters. The information we wanted to expose to CK had to be implemented in \texttt{cl\_launcher}. The process itself was fairly simple, as we knew how to obtain the data we wanted to record, and, most importantly, does not modify the normal functionality of the program.

\paragraph{CK integration discussion}
The two testing processes are sketched in Figures \ref{fig:testinginit} and \ref{fig:testingck}. While functionally they do the same job, the issues of maintenance and accessibility are on a different level. In our initial testing infrastructure, we developed multiple versions of our scripts, having the same code appear multiple times throughout them and generally doing a bad job of keeping track of everything in local and remote repositories. In addition, we did not even consider documenting these, preferring to guide users through the whole process. On the other hand, the CK integration abstracts most of this away, exposing only the information pertinent to our project, through \emph{programs} and \emph{datasets}. Since it offers a unified pipeline, we can simply take it as is, assured that previous project applications have tested it thoroughly. Not to mention that having CK manage data parsing is a tremendous improvement compared to having to do it manually.

We are uncertain about the technical difficulty of integrating a project in CK. The CLsmith project required roughly half a day of work from an expert in order to be integrated, followed by a few hours of configuring the view according to our needs. The modifications on our code were minimal, but did seem to require expert knowledge of the framework. Also, we had a single executable which required integration. As such, we cannot comment on how feasible it would be to use the framework for a bigger project, with multiple executables, with only amateur knowledge.

The current existing infrastructure uses a number of scripts to abstract away the required knowledge the user is expected to have. Unfortunately, the underlying mechanisms are hand crafted JSON files; in order to modify them or create new possible tests, these have to be analysed and understood. This might pose an issue, as their structure is quite intimate with the inner workings of CK.

\section{Future Work}
The next step to be done in this collaboration is to integrate the entire CLsmith testing process into CK. This means also adding the generator, implementing the ability to generate tests on the fly and having the ability to analyse results macroscopically, including majority voting and bug classification. These features would require extra changes in what CK records, as we also need to record what version of the generator was used for a particular test. This means we are required to also record which version of the generator was used for a specific program and to limit our analysis to that particular generation. Majority voting is also beyond the scope of CK, requiring implementation by a script. However, seeing as the results and tests are handled by the framework, we believe this to be a degree easier than our ad hoc implementation, where we also had to gather and parse the results ourselves.

We would also like to integrate the possibility of uploading reduced versions of the existing tests in the CK repository. This would encourage community collaboration, something the CK developers are aiming for. A requirement for this would be the ability to record the original program the reduction comes from, as well as the nature of the bug being investigated. As a consequence, we would prefer to have a straightforward method of tailoring the experimental view to our needs. Currently, we believe this also required expert knowledge of CK.

Aside from allowing the opportunity of sharing information via manual reduction, we envision the possibility of integrating an automated reduction process that could be done for newly uploaded CLsmith generated kernels. Some work has been done as an extension of the C-Reduce tool \cite{creduce} lifted to the OpenCL language \cite{clreduce}, but we aren't certain if the current status of CK would allow for such an involved activity. We believe this would allow for researchers uninitialised in how CLsmith or OpenCL work to make use of the generator and the bug-finding mechanism in a simple way, abstracting all the inner workings of the project via CK. However, this requires both projects to mature some more.

\section*{Acknowledgements}
We would like to thank Grigori Fursin and Anton Lokhmotov for bringing CK to our attention, their great help with integrating CLsmith into CK and generally making this paper possible.

\bibliographystyle{abbrv}
\bibliography{references.bib}

\begin{thebibliography}{10}

\bibitem{ckguide}
G.~Fursin and A.~Lokhmotov.
\newblock Getting started with {CK}.
\newblock \url{https://github.com/ctuning/ck/wiki/Getting_started_guide}, 2015.

\bibitem{ckclsmith}
G.~Fursin and A.~Lokhmotov.
\newblock Getting started with {CLsmith} in {CK} format.
\newblock
  \url{https://github.com/ctuning/ck/wiki/Getting_started_guide_clsmith}, 2015.

\bibitem{collectivemind2}
G.~{Fursin}, A.~{Memon}, C.~{Guillon}, and A.~{Lokhmotov}.
\newblock {Collective Mind, Part II: Towards Performance- and Cost-Aware
  Software Engineering as a Natural Science}.
\newblock {\em ArXiv e-prints}, June 2015.

\bibitem{collectivemind}
G.~Fursin, R.~Miceli, A.~Lokhmotov, M.~Gerndt, M.~Baboulin, D.~Malony, Allen,
  Z.~Chamski, D.~Novillo, and D.~D. Vento.
\newblock {Collective mind: Towards practical and collaborative auto-tuning}.
\newblock {\em Scientific Programming}, 22(4):309--329, July 2014.

\bibitem{opencl12}
{Khronos}.
\newblock The {OpenCL} specification, version 1.2, 2012.
\newblock Document revision 19.

\bibitem{emi}
V.~Le, M.~Afshari, and Z.~Su.
\newblock Compiler validation via equivalence modulo inputs.
\newblock In {\em {ACM} {SIGPLAN} Conference on Programming Language Design and
  Implementation, {PLDI} '14, Edinburgh, United Kingdom - June 09 - 11, 2014},
  page~25, 2014.

\bibitem{clsmith}
C.~Lidbury, A.~Lascu, N.~Chong, and A.~F. Donaldson.
\newblock Many-core compiler fuzzing.
\newblock In {\em Proceedings of the 36th ACM SIGPLAN Conference on Programming
  Language Design and Implementation}, PLDI 2015, pages 65--76, New York, NY,
  USA, 2015. ACM.

\bibitem{clreduce}
P.~Moritz.
\newblock Automatic test case reduction of randomly generated {OpenCL} kernels.
\newblock Master's thesis, Imperial College London, London, UK, 2015.

\bibitem{creduce}
J.~Regehr, Y.~Chen, P.~Cuoq, E.~Eide, C.~Ellison, and X.~Yang.
\newblock Test-case reduction for {C} compiler bugs.
\newblock In {\em {ACM} {SIGPLAN} Conference on Programming Language Design and
  Implementation, {PLDI} '12, Beijing, China - June 11 - 16, 2012}, pages
  335--346, 2012.

\bibitem{scikit-learn}
G.~Varoquaux, L.~Buitinck, G.~Louppe, O.~Grisel, F.~Pedregosa, and A.~Mueller.
\newblock Scikit-learn: Machine learning without learning the machinery.
\newblock {\em GetMobile}, 19(1):29--33, 2015.

\bibitem{csmith}
X.~Yang, Y.~Chen, E.~Eide, and J.~Regehr.
\newblock Finding and understanding bugs in {C} compilers.
\newblock In {\em Proceedings of the 32nd {ACM} {SIGPLAN} Conference on
  Programming Language Design and Implementation, {PLDI} 2011, San Jose, CA,
  USA, June 4-8, 2011}, pages 283--294, 2011.

\end{thebibliography}

\end{document}